\documentclass[twocolumn]{aastex631} 

\usepackage{float} 
\usepackage{amsmath}

\begin{document}

\title{Central Concentration and Escape of Ionizing Photons in Galaxies at the Epoch of Reionization}

\correspondingauthor{Cheqiu Lyu, Enci Wang}
\email{lyucq@ustc.edu.cn, ecwang16@ustc.edu.cn}

\author[0009-0000-7307-6362]{Cheqiu Lyu}
\affiliation{Department of Astronomy, University of Science and Technology of China, Hefei, Anhui 230026, China}
\affiliation{School of Astronomy and Space Science, University of Science and Technology of China, Hefei, Anhui 230026, China}

\author[0000-0003-1588-9394]{Enci Wang}
\affiliation{Department of Astronomy, University of Science and Technology of China, Hefei, Anhui 230026, China}
\affiliation{School of Astronomy and Space Science, University of Science and Technology of China, Hefei, Anhui 230026, China}

\author[0000-0002-4419-6434]{Junxian Wang}
\affiliation{Department of Astronomy, University of Science and Technology of China, Hefei, Anhui 230026, China}
\affiliation{School of Astronomy and Space Science, University of Science and Technology of China, Hefei, Anhui 230026, China}

\author[0009-0004-7042-4172]{Cheng Jia}
\affiliation{Department of Astronomy, University of Science and Technology of China, Hefei, Anhui 230026, China}
\affiliation{School of Astronomy and Space Science, University of Science and Technology of China, Hefei, Anhui 230026, China}

\author[0000-0002-0846-7591]{Jie Song}
\affiliation{Department of Astronomy, University of Science and Technology of China, Hefei, Anhui 230026, China}
\affiliation{School of Astronomy and Space Science, University of Science and Technology of China, Hefei, Anhui 230026, China}

\author[0000-0002-4597-5798]{Yangyao Chen}
\affiliation{Department of Astronomy, University of Science and Technology of China, Hefei, Anhui 230026, China}
\affiliation{School of Astronomy and Space Science, University of Science and Technology of China, Hefei, Anhui 230026, China}

\author[0009-0004-5989-6005]{Zeyu Chen}
\affiliation{Department of Astronomy, University of Science and Technology of China, Hefei, Anhui 230026, China}
\affiliation{School of Astronomy and Space Science, University of Science and Technology of China, Hefei, Anhui 230026, China}

\author[0009-0008-1319-498X]{Haoran Yu}
\affiliation{Department of Astronomy, University of Science and Technology of China, Hefei, Anhui 230026, China}
\affiliation{School of Astronomy and Space Science, University of Science and Technology of China, Hefei, Anhui 230026, China}

\author[0009-0006-7343-8013]{Chengyu Ma}
\affiliation{Department of Astronomy, University of Science and Technology of China, Hefei, Anhui 230026, China}
\affiliation{School of Astronomy and Space Science, University of Science and Technology of China, Hefei, Anhui 230026, China}

\author{Jinyang Wang}
\affiliation{Department of Astronomy, University of Science and Technology of China, Hefei, Anhui 230026, China}
\affiliation{School of Astronomy and Space Science, University of Science and Technology of China, Hefei, Anhui 230026, China}

\author{Yifan Wang}
\affiliation{Department of Astronomy, University of Science and Technology of China, Hefei, Anhui 230026, China}
\affiliation{School of Astronomy and Space Science, University of Science and Technology of China, Hefei, Anhui 230026, China}

\author[0000-0002-7660-2273]{Xu Kong}
\affiliation{Department of Astronomy, University of Science and Technology of China, Hefei, Anhui 230026, China}
\affiliation{School of Astronomy and Space Science, University of Science and Technology of China, Hefei, Anhui 230026, China}


\begin{abstract}
Compact, low-mass galaxies with strong nebular emission are considered promising candidates for efficient ionizing photon production and escape. We present a spatially resolved analysis of 189 galaxies at redshifts $z \sim 6.7-7.6$ in JADES GOODS-N and GOODS-S fields and selected via JWST/NIRCam F410M filter. By employing annular photometry and spectral energy distribution fitting across rest-frame UV to optical wavelengths, we investigate the internal structure of star formation, ionizing photon production and escape, as well as the resolved star formation histories within these galaxies. We find that these galaxies exhibit compact, centrally concentrated, and bursty star formation, especially in lower-mass systems ($\log M_*/{\rm M_{\odot}} <9.0$). The central regions of them display extreme [OIII]+H$\beta$ equivalent widths ($>$1000 \AA), high ionizing photon production efficiencies ($\xi_{\text{ion}} \sim 10^{25.6}$ Hz erg$^{-1}$), steep UV slopes ($\sim -2.3$), and elevated escape fractions ($f_{\text{esc}} > 0.08$), with all these properties peaking in the inner regions. These findings reveal outside-in growth and rising star formation histories at $z\sim 7$, with the central regions of them playing a pivotal role in driving cosmic reionization.

\end{abstract}


\section{Introduction} \label{sec:intro}

The Epoch of Reionization (EoR) marks a pivotal transition in cosmic history, when the first generation of stars and galaxies formed and began ionizing the surrounding neutral hydrogen. This process allowed Lyman Continuum (LyC) photons to propagate through the intergalactic medium (IGM), dramatically transforming the thermal and ionization state of the Universe \citep[e.g.,][]{Bromm2011, Robertson2015, Bosman2022, Robertson2023, Stark2025}. Observations constrain the end of this epoch to redshifts $z \sim 6$ \citep[e.g.,][]{Becker2001, Fan2006, Yang2020}. While there is consensus that young, massive stars are key producers of ionizing photons capable of escaping the interstellar medium (ISM) and ionizing the IGM \citep[e.g.,][]{Hassan2018}, the nature of the dominant reionizing sources—whether bright and massive or faint and numerous low-mass galaxies—remains under debate \citep{Finkelstein2019, Naidu2020, Robertson2022, Yeh2023}.

During the EoR, some galaxies are expected to undergo intense, bursty episodes of star formation, making them key contributors to the cosmic ionizing photon budget. These systems are characterized by very young stellar populations, resulting in extremely faint rest-frame optical continua and correspondingly large equivalent widths (EWs) of nebular emission lines, including [OIII], H$\beta$, and H$\alpha$. The dominance of line emission over continuum flux leads to noticeable flux excesses in broadband or medium-band filters. Such excesses, particularly in filters covering the [OIII]+H$\beta$ complex, serve as a powerful observational signature for identifying galaxies with high specific star formation rates (sSFR) and intense ongoing star formation \citep{Stark2025}. Thanks to the medium-band filters of JWST and the sensitivity of NIRCam imaging \citep{Gardner2023}, these features can now be detected out to the epoch of reionization, enabling the efficient selection of line-excess galaxies at $z > 6$ \citep[e.g.,][]{Endsley2024, Begley2025}.

Among the population of strongly star-forming galaxies, [OIII]-excess systems—characterized by unusually strong [OIII] emission—are particularly compelling candidates for contributing significantly to cosmic reionization. These galaxies typically exhibit high star formation rates (SFRs), extreme ionization conditions, and low gas-phase metallicities, all of which enhance the production and potential escape of ionizing photons. Two key parameters govern a galaxy’s ability to influence reionization: the ionizing photon production efficiency ($\xi_{\mathrm{ion}}$), and the escape fraction of those photons into the intergalactic medium ($f_{\mathrm{esc}}$) \citep{Robertson2015, Finkelstein2019, Mason2019, Begley2025, Stark2025}.

Observationally, many confirmed LyC-leaking galaxies show strong [OIII]$\lambda5007$ and H$\beta$ emission, often accompanied by high O32 ratios ([OIII]$\lambda\lambda$4959,5007/[OII]$\lambda\lambda$3726,3729), which are indicative of hard ionizing spectra and low column densities \citep{Vanzella2016, Izotov2018, Fletcher2019, Izotov2021}. These line diagnostics provide valuable insight into the ionization state and geometry of the ISM. Furthermore, galaxies with extreme [OIII] emission tend to exhibit elevated $\xi_{\mathrm{ion}}$ values, suggesting that [OIII]-excess systems may serve as efficient tracers of high LyC output and leakage \citep{Chevallard2018, Tang2019, Onodera2020}. Consequently, identifying and characterizing such galaxies is essential to advancing our understanding of the sources that reionized the early Universe.

While early EoR galaxy studies have predominantly focused on integrated properties, such measurements can obscure the diversity of internal physical conditions. Spatially resolved observations provide critical insights into the structure and distribution of star formation and gas properties within galaxies. High-resolution JWST/NIRCam imaging has opened a new window into the internal dynamics and morphology of EoR galaxies, offering unprecedented clarity on how local processes scale up to influence global reionization. For instance, \citet{Tripodi2024} analyzed spatially resolved emission and H$\beta$ EW gradients for 63 galaxies at $4 \leq z < 10$ using JWST/NIRSpec data, finding negative radial EW gradients indicative of centrally concentrated starbursts fueled by high accretion rates. In this work, we aim to build on this framework by investigating the spatially resolved properties of galaxies during the EoR, leveraging deep imaging from the JADES survey. By examining the internal structure and spatial distribution of [OIII] emission-related properties, we seek to elucidate the mechanisms driving the [OIII] emission in these galaxies and their broader implications for the reionization process. This paper is organized as follows: Section \ref{sec:data} describes the parent image and catalog, and sample selection. Section \ref{sec:processing} describes the method of photometry measurements and SED fitting. In Section \ref{sec:results}, we analyze and discuss the derived integrated physical properties and their profiles. Section \ref{sec:summary} summarizes our findings and gives conclusions. Throughout this work, we assume a flat cold dark matter cosmology model with $\Omega_m=0.3, \Omega_\Lambda=0.7$, and $h=0.7$ when computing distance-dependent parameters.

\section{data and sample selection} \label{sec:data}

\subsection{DJA Images and Catalogs}
The initial dataset utilized in this work originates from the JWST Advanced Deep Extragalactic Survey (JADES) Data Release Version 2.0 \citep{Eisenstein2023a, Eisenstein2023b, Rieke2023}, which includes observations across nine JWST/NIRCam filters: F090W, F115W, F150W, F200W, F277W, F335M, F356W, F410M, and F444W. With exposure times exceeding 10 hours per filter, JADES delivers the deepest near-infrared imaging to date, reaching a 5$\sigma$ limiting depth of $\sim$ 30 AB magnitudes. The survey covers approximately 60 $\mathrm{arcmin}^{2}$ across the GOODS-North (GOODS-N) and GOODS-South (GOODS-S) fields \citep{Brandt2001, Giacconi2002, Giavalisco2004}.

In this work, we utilize the NIRCam imaging mosaics (v7.2 and v7.3) from the Cosmic Dawn Center JWST Archive (DJA). This public repository of JWST galaxy data is reduced with \texttt{grizli} \citep{Brammer2019} and \texttt{msaexp} \citep{Brammer2022}. Our photometric parent catalog is the DJA Morphological Catalog (hereafter DJA catalog) \citep{Bertin1996}, constructed using \texttt{SourceExtractor++} and available online\footnote{\url{https://dawn-cph.github.io/dja/blog/2024/08/16/morphological-data/}}. It includes 70,421 sources in the GOODS-N field and 70,357 sources in the GOODS-S field, with detailed model-based morphological measurements, fixed-aperture photometry, and derived physical properties such as stellar masses, SFRs, and stellar population parameters. Photometric redshifts ($z_\mathrm{phot}$) are computed using \texttt{EAZY-PY} \citep{Brammer2008}, with normalized median absolute deviations ($\sigma_{\mathrm{NMAD}}$) of 0.019 and 0.016 for the JADES-GDS and JADES-GDN fields, respectively \citep{Valentino2023}, where $\sigma_{\mathrm{NMAD}}$ is defined as the median absolute deviation of $\Delta z/(1+z_\mathrm{spec})$. Basic details of the NIRCam data reduction are presented in \citet{Valentino2023}.

\subsection{Sample Selection}

\begin{figure*}[htpb!]
    \centering
    \includegraphics[width=1\linewidth]{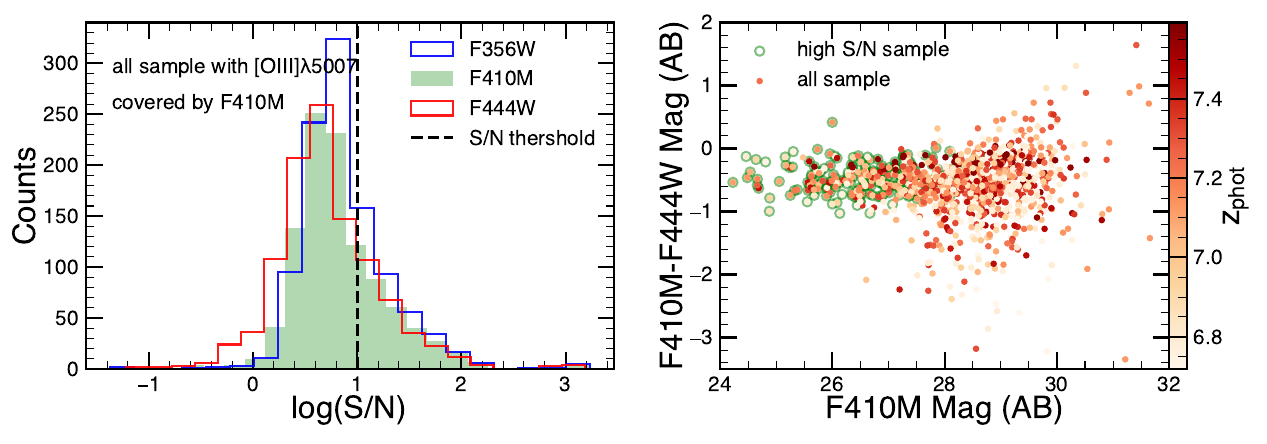}
    
    \caption{Data selection and classification. Left: Signal-to-noise ratio (S/N) distributions in the F356W, F410M, and F444W filters for all galaxies with [OIII]$\lambda$5007 emission covered by the F410M filter. The black dashed line marks the adopted S/N threshold. Right: Color–magnitude diagram color-coded by photometric redshift ($z_{\mathrm{phot}}$). Green-edged points indicate sources that meet the S/N selection criterion.}
    \label{DataSelection}
\end{figure*}

We select the sources with stellar masses exceeding $10^7\mathrm{M_\odot}$ and reliable photometry in the medium-band filter F410M as well as adjacent broad-band filters F356W and F444W. The F410M band captures the [OIII]+H$\beta$ emission complex, including the [OIII]$\lambda\lambda$4959,5007 and H$\beta$ ($\lambda$4861) lines for galaxies at $z \sim 6.7$–$7.6$. To ensure robust redshift estimates, we further discard galaxies exhibiting large $z_{\mathrm{phot}}$ uncertainties, specifically those where $(z_{84} - z_{16})/2 > 1$. Here, $z_{84}$ and $z_{16}$ denote the 84th and 16th percentiles of the posterior distribution. We utilize the stellar mass values from the DJA catalog for the initial data selection. In the subsequent analyses presented in this paper, we employ the stellar mass derived from our SED fitting, which is consistent with those from the DJA catalog, with a mean offset of approximately 0.07 dex.

The left panel of Figure~\ref{DataSelection} presents the S/N distributions in F356W, F410M, and F444W of the cutout images for candidate galaxies. To ensure the robustness of spatially resolved photometric and SED analyses, we select high signal-to-noise (S/N) sources by adopting a conservative threshold of S/N $\geq$ 10 in all three bands. The right panel of Figure~\ref{DataSelection} shows the  F410M–F444W color–magnitude diagram with highlighting selected sources with green-edged markers. As shown, most of the selected galaxies are brighter in F410M than F444W (F410M-F444W$<$0), especially for F410M-brighter ones. This step results in a sample of 217 galaxies. For all candidates, we visually inspect the multi-band JWST cutouts provided in the DJA catalog and exclude 8 sources with possible contamination or artifacts that could bias photometry or SED fitting. After applying these selection criteria, we obtain a sample comprising 209 galaxies at $z \sim 6.7$–$7.6$ from the JADES GOODS-N and GOODS-S fields. Of these, 189 galaxies were successfully fitted with SED models (Section \ref{SEDfitting}) and constitute the sample used in the following analysis.

It is important to note that, for spatially resolved statistical studies, we impose an S/N threshold in the F356W, F410M, and F444M bands when selecting sources. This inevitably led to the rejection of some fainter samples in the broad band, whose properties might be more extreme. These will be investigated in our future work.

\section{data processing}\label{sec:processing}

\subsection{Photometry Measurement}
To construct multi-wavelength photometric catalogs, we resample the short-wavelength filters (F090W, F115W, F150W, F200W) images to match the resolution of the long-wavelength filters (F277W, F335M, F356W, F410M, F444W), ensuring a uniform pixel scale of 0.04 arcsec per pixel. In addition, we homogenize the point spread functions (PSFs) across different filters by convolving each image to match the F444W PSF \citep[e.g.,][]{McLeod2024}. The convolution kernels are derived from empirical PSF models in the DJA \texttt{gds-grizli-v7.2} and \texttt{gdn-grizli-v7.3} datasets for the GOODS-N and GOODS-S fields, respectively. This ensures consistent image resolution, allowing for a comparative analysis of galaxy properties across various rest-frame wavelengths. The full width at half maximum (FWHM) of the PSF in F444W is about $2.10$ pixel (0.084 aresec).

We generate 1.2 arcsec × 1.2 arcsec cutouts (corresponding to 30 × 30 NIRCam pixels) for each galaxy in all nine filters. The center coordinates of each cutout image are provided in the DJA catalog. To measure the total and spatially resolved flux of each galaxy, we perform photometry using \texttt{Photutils} \citep{Bradley2022}. Circular and annular apertures are centered on the detected positions of sources in each image. For each galaxy, circular apertures are set to a fixed radius of 12 pixels (0.48 arcsec), and the annular apertures ranges are 0-1, 1-2, 2-3, 3-4, 4-5, 5-6, 6-7, 7-8, 8-10, 10-12 pixel radii (see example in Figure \ref{SEDExample}). The stacked cutout images and surface profiles for the F356W, F410M, and F444W bands, and composite RGB image for these galaxies are shown in Figure \ref{stack} in Appendix \ref{app1}.

\subsection{SED Fitting}\label{SEDfitting}
To derive the best-fitting spectral energy distribution (SED) models and physical properties for each galaxy and its regions, we perform multi-wavelength SED fitting using the Bayesian SED modeling code \texttt{BAGPIPES} \citep[Bayesian Analysis of Galaxies for Physical Inference and Parameter Estimations,][]{Carnall2018, Carnall2019}. This method allows for self-consistent estimation of key physical parameters such as stellar mass, star formation history (SFH), and age.

We employ broad, uniform priors in our model: stellar mass is allowed to vary between $10^{5}$ and $10^{12}$$\mathrm{M_{\odot}}$, while stellar metallicity is constrained to the range [0.01, 5]. Dust attenuation is modeled using the Calzetti dust law \citep{Calzetti2010}, with $A_v$ values ranging from 0 to 3. We incorporate emission lines and nebular continuum based on  \texttt{Cloudy} \citep{Ferland2013}. To explore a broader range of ionization parameters, we regenerate \texttt{Cloudy} models using a configuration file distributed with \texttt{BAGPIPES}, covering the ionization parameter $U$ varying between $-4$ and $-1$. The free parameters used in the fiducial \texttt{BAGPIPES} models for this work are similar to those in \citet{Begley2025}, who explored a range of model fitting configurations for galaxies at $z = 6.9–7.6$, varying stellar population synthesis (SPS) templates, SFHs, and dust attenuation prescriptions. Specifically, we adopt a non-parametric `continuity' SFH \citep[e.g.,][]{Leja2019}, which is modeled using the BPASS SPS models \citep{Eldridge2017}. The continuity SFH is divided into five time bins: 0–3 Myr, 3–10 Myr, 10–100 Myr, 100–250 Myr, and 250–500 Myr, which is consistent with the time bins used in recent literature \citep[e.g.,][]{Whitler2023, Endsley2024, Begley2025}. Given the photometric redshift selection of the JWST sample, we fit each galaxy and region with a fixed redshift prior of $z_{\text{phot}}$.

\begin{figure*}[htpb!]
    \centering
    \includegraphics[width=1\linewidth]{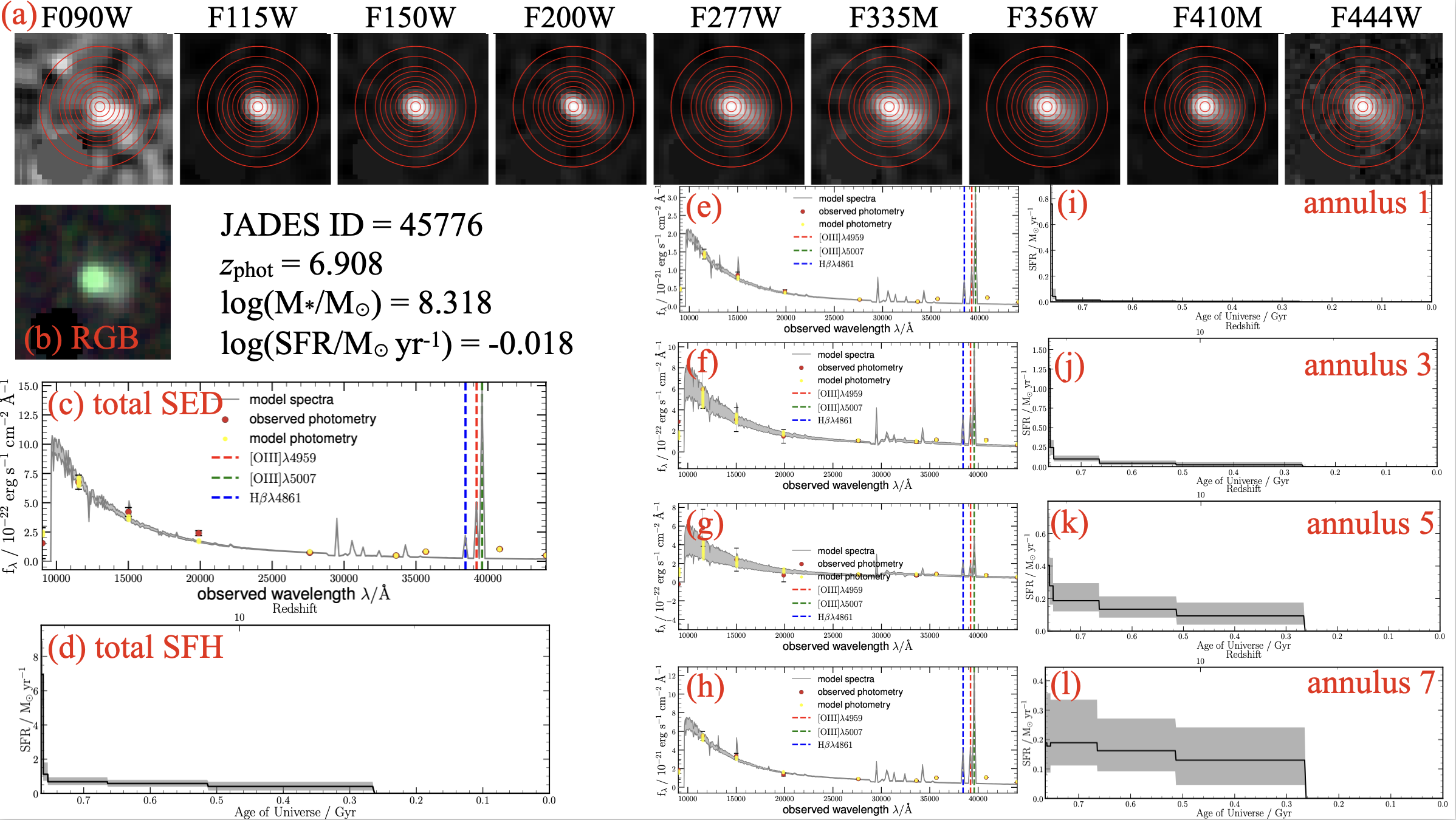}
    
    \caption{Photometric measurements and spectral energy distribution fitting for a representative galaxy (ID 45776) at $z_{\mathrm{phot}} = 6.908$ in the JADES GOODS-S field. Panel (a): Multi-band cutouts convolved to match the F444W PSF, with red circles marking the photometric annuli. Panel (b): RGB color composite image constructed from the F444W (R), F410M (G), and F356W (B) filters. The pixel resolution and cutout scale is 0.04 and 1.2 arcsec, corresponding to $\sim 0.21$ and $\sim 6.3$ kpc at the target redshift, respectively. Panels (c) and (d): SED and 500 fitting model spectra generated with \texttt{BAGPIPES}, along with the corresponding star formation history. Panels (e)–(h): Spatially resolved SEDs and fitted spectra for concentric annuli at pixel radii [0–1], [2–3], [4–5], and [6–7], respectively. Panels (i)–(l): Star formation histories associated with the annuli shown in panels (e)–(h).}
    \label{SEDExample}
\end{figure*}

For the SED fitting, we only include galaxies and regions with S/N $>$ 5 in the F410M filter. For bands where a galaxy/region is undetected, we include those filters (typically no more than one filter) in the fitting by assigning a flux value of zero along with a large associated uncertainty. For each galaxy or region, the SED fitting generates 500 posterior model spectra and corresponding photometry across the nine bands. An example of the best-fitting posterior SED models and SFHs for a whole galaxy and a few regions is shown in Figure \ref{SEDExample}.

\section{Results and discussion} \label{sec:results}

\subsection{Star-Forming Main Sequence}
\begin{figure}[htpb!]
    \centering
    \includegraphics[width=1\linewidth]{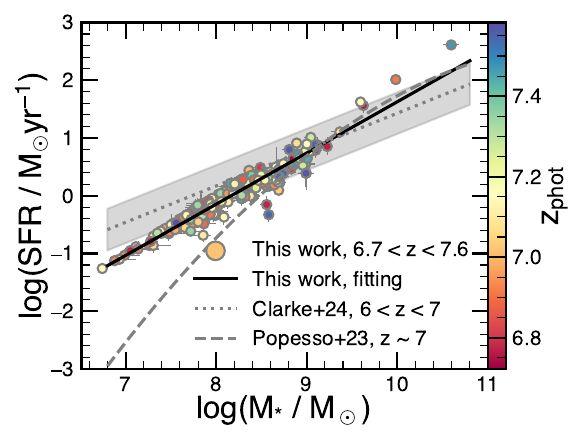}
    
    \caption{Stellar mass–SFR relation for our sample. Each point is color-coded by the galaxy’s photometric redshift ($z_{\mathrm{phot}}$). Short grey lines indicate the 16th and 84th percentile ranges. The solid black line shows the best-fit linear relation for our sample. For comparison, the grey dotted and dashed lines represent empirical main sequences of star-forming galaxies at $6 < z < 7$ from \citet{Clarke2024} and at $z \sim 7$ from \citet{Popesso2023}, respectively.}
    \label{SFMS}
\end{figure}

The stellar mass and SFR of each galaxy are determined by the median values from fitting 500 posterior spectra using \texttt{BAGPIPES}, with the derived SFR corresponding to a 100 Myr timescale, as \texttt{BAGPIPES} default setting. As shown in Figure \ref{SFMS}, our sample follows a relatively tight star-forming main sequence (SFMS). The best-fit linear relation for our sample is:
\begin{equation}
\log(\mathrm{SFR}/\mathrm{M}_{\odot} \mathrm{yr}^{-1}) = 0.887 \log(M_\mathrm{*}/\ M_\odot) - 7.24,
\end{equation}
which aligns well with similar redshift samples in the literature \citep[e.g.,][]{Popesso2023, Clarke2024}. 

We divide our galaxy sample into three stellar mass bins for a more detailed analysis of mass-dependent galaxy properties: $\log(M_*/\mathrm{M}_\odot) = 6.5-8.0$ (M1: [6.5,8], 71 galaxies), $8.0-9.0$ (M2: [8,9], 101 galaxies), and $9.0-11.0$ (M3: [9,11], 17 galaxies).

\subsection{UV Continuum Slope and [OIII]+H$\beta$ Equivalent Width}

\begin{figure*}[htpb!]
    \centering
    \includegraphics[width=1\linewidth]{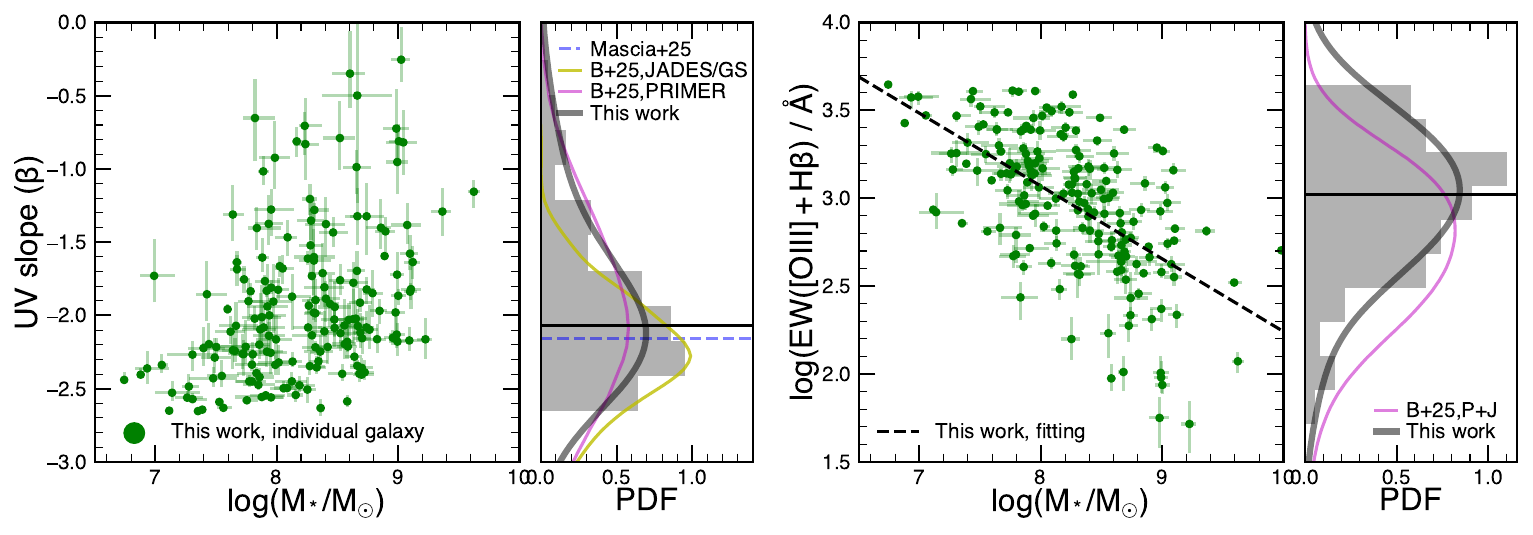}
    
    \caption{Left: Total UV continuum slope ($\beta$) as a function of stellar mass for our galaxy samples. Short lines denote the 16th and 84th percentile ranges for each galaxy. The side panel presents the $\beta$ distribution along with kernel density estimation (KDE) curves. The blue dashed line marks the median value for galaxies at $5 < z < 7$ from \citet[][Mascia+25]{Mascia2025}. Yellow and magenta curves show KDE distributions from the JADES/GOODS-S and PRIMER samples in \citet[][B+25]{Begley2025}. Right: Rest-frame [OIII]+H$\beta$ equivalent width as a function of stellar mass. The black dashed line represents the linear fit of our sample. The symbol coding and histogram format follow those of the left panel. The magenta KDE curve represents the PRIMER+JADES sample from \citet[][B+25]{Begley2025}.}
    \label{UVEW}
\end{figure*}
The UV continuum slope, $\beta$, where $f_{\lambda} \propto \lambda^{\beta}$, is a crucial indicator of high-redshift galaxy properties linked to the contribution to reionization. Blue UV slopes (lower $\beta$) typically suggest that these galaxies are young, have relatively low metallicity, and low levels of dust attenuation or even dust-free \citep[e.g.,][]{Cullen2023, Cullen2024}. These blue slopes imply higher-than-average ionizing photon production efficiencies \citep{Cullen2024, Topping2024}, and are often associated with significant ionizing photon escape fractions ($f_{\mathrm{esc}}$) \citep{Begley2022, Chisholm2022, Choustikov2024, Kreilgaard2024}.

For each posterior spectrum, we fit a power law to the UV continuum in the rest-frame range of 1300\AA\ to 1800\AA\ to obtain $\beta$  \citep{Chisholm2022}. The UV slope for each galaxy and region is calculated as the median value from fitting 500 posterior spectra.

The left panel of Figure \ref{UVEW} shows that the UV slope positively correlates with the stellar mass of the galaxy sample. The average UV slope for our sample is $\beta=-2.07$ with a standard deviation of 0.51, with the black solid line representing the median. This distribution is consistent with recent measurements of UV slopes for galaxies at $z \sim 5$-7 \citep[e.g.,][]{Begley2025, Mascia2025}.

The composite color images of our sample reveal resolved compact green components, which represent regions of strong [OIII]+H$\beta$ emission with equivalent widths exceeding several hundred angstroms \citep{Cardamone2009}. For each galaxy, we use the median flux in the F410M filter and posterior spectra to estimate the [OIII]+H$\beta$ equivalent width, EW([OIII]+H$\beta$). We assume that the emission lines in F410M are dominated by [OIII]+H$\beta$. Specifically, we calculate the continuum flux by fitting a power-law continuum in the 4500-4800\AA\ and 5100-5300\AA\ windows, then subtract this continuum from the F410M flux. The equivalent width is given by: 
\begin{equation}
\mathrm{EW([OIII]+H\beta)} = \frac{f_\mathrm{{F410M,\lambda}}  - f_\mathrm{{con,\lambda}} }{f_\mathrm{{con,\lambda}}} \times \Delta W/(1+z),
\end{equation}

where the $f_\mathrm{{F410M,\lambda}}$ and $f_\mathrm{{con,\lambda}}$ indicates the flux of F410M band and continuum, respectively, in the unit of $\mathrm{erg\ s^{-1}\ cm^{-2}\ \AA^{-1}}$. The photometric flux unit of $\mu$Jy has been converted to $\mathrm{erg\ s^{-1}\ cm^{-2}\ \AA^{-1}}$ using the pivot wavelength of $4.083\ \mu$m for the F410M band. Additionally, $\Delta W$ represents the bandwidth of $0.436\ \mu$m for the F410M band\footnote{\url{https://jwst-docs.stsci.edu/jwst-near-infrared-camera/nircam-instrumentation/nircam-filter}}. The EW([OIII]+H$\beta$) for each galaxy is derived from the median of 500 posterior spectra.

The right panel of Figure \ref{UVEW} shows the EW([OIII]+H$\beta$) negatively correlates with stellar mass for the galaxy sample, with the best-fit linear relation:
\begin{equation}
\mathrm{\log (EW([OIII]+H\beta)/ \AA )} = -0.414 \log(M_*/{\mathrm{M_\odot}}) + 6.38,
\end{equation}
The average EW([OIII]+H$\beta$) for our entire sample is 1000\AA\ with a standard deviation of 0.39 dex. This distribution is consistent with recent measurements of galaxies at $z \sim 7$ \citep[e.g.,][]{Begley2025}.

\subsection{Star Formation Profiles}

\begin{figure*}[htpb!]
    \centering
    \includegraphics[width=1\linewidth]{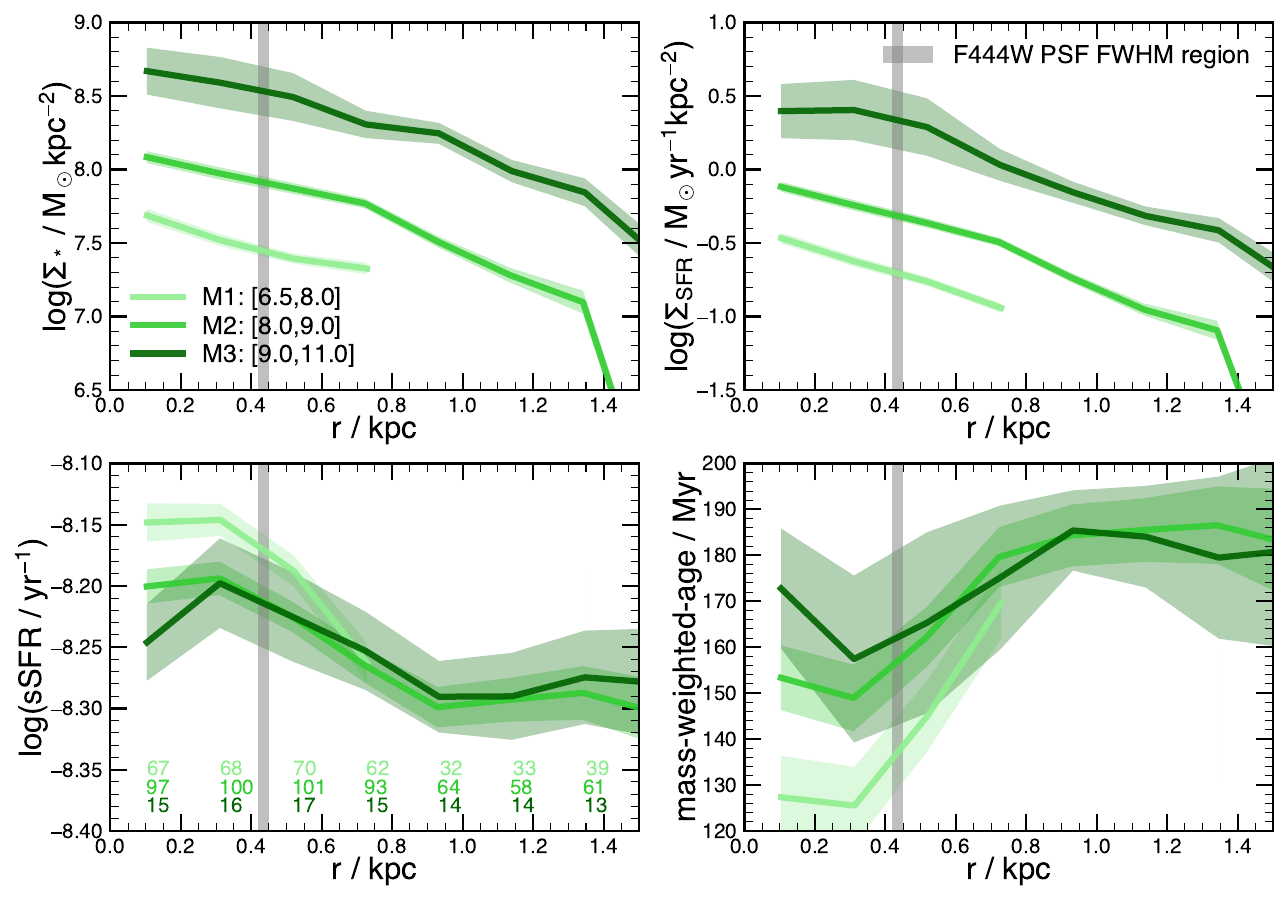}
    
    \caption{Radial profiles of stellar mass projected surface density ($\Sigma_*$, top left), SFR projected surface density ($\Sigma_{\mathrm{SFR}}$, top right), sSFR (bottom left), and mass-weighted age (bottom right). The light-green, green, and dark-green curves represent the median profiles of galaxies in three stellar mass bins. Shaded regions represent the $1\sigma$ uncertainties, calculated as $1.253 \times \sigma / \sqrt{n}$, where $\sigma$ is the standard deviation of the valid data points within each radial bin, and $n$ is the number of valid data points in that bin, also indicated at the bottom of the bottom-left panel. The grey shaded area indicates the range of the F444W PSF FWHM across the redshift range of our sample.}
    \label{SDprofile}
\end{figure*}

Figure \ref{SDprofile} presents the radial distributions of key star formation-related properties for our galaxy sample across different stellar mass bins. These properties include stellar mass surface density ($\Sigma_*$, top left), star formation rate surface density ($\Sigma_{\mathrm{SFR}}$, top right), specific star formation rate (sSFR, bottom left), and mass-weighted stellar age (bottom right). On average, both $\Sigma_*$ and $\Sigma_{\mathrm{SFR}}$ exhibit negative gradients, indicative of centrally concentrated stellar mass and star formation activity. Higher-mass galaxies show systematically higher profiles, suggesting more extended and sustained star formation activity. 

We initially perform annular photometric measurements on the DJA images at the pixel level and subsequently convert pixel scales into physical units (kpc) for further analysis. At $z \sim$ 6.7-7.6, one pixel (0.04 arcsec) corresponds to a physical scale of approximately 0.199–0.214 kpc. It is worth noting that measurements within the innermost $\sim$ 0.4 kpc may be affected by PSF smearing and resolution limitations, and should be interpreted with caution. The radial profiles normalized by effective radius are presented in Appendix \ref{app2}.

To minimize artificial radial trends caused by missing SED fitting results, we assign NaN values to regions without reliable fits. In each radial bin, if fewer than half of the data points are NaN, we compute the median by including all regions for $\Sigma_{*}$ and $\Sigma_{\rm SFR}$ profiles, treating NaNs as lower limits, indicative of values below the detection threshold (e.g., low-S/N regions). Test show that this treatment does not significantly change the overall results compared to simply excluding NaNs. For radial profiles of sSFR, mass-weighted age, EW([OIII]+H$\beta$), and UV slope, median values are calculated using only valid measurements. This approach introduces significant sample incompleteness at large radii. We therefore exclude radial bins where $>$50\% of values are NaN. We also indicate the number of valid fitting regions in each radial bin in the plots for all stellar mass bins. Annular photometry has been performed in 10 radial apertures, extending to 0.48" ($\sim$ 2.4 kpc). However, due to increasing incompleteness beyond 1.5 kpc, we limit the x-axis to within 1.5 kpc in the final profiles.

The bottom panels of Figure \ref{SDprofile} show that sSFR declines with radius, while the mass-weighted stellar age increases outward within approximately 1 kpc. These profiles tend to flatten at $r>$ 1kpc, especially in high-mass galaxies, which may be due to the poor statistics of insufficient valid fittings. Notably, these trends within 1 kpc indicate that the central regions host younger stellar populations and exhibit higher sSFRs, whereas the outer regions formed earlier. This pattern suggests an outside-in mode of stellar mass assembly at these early epochs—markedly contrasting with the well-established inside-out growth scenario observed at lower and intermediate redshifts, where galactic centers are typically older than their outskirts \citep[e.g.,][]{Wuyts2013, Li2015, Ibarra-Medel-16, Goddard-17, Wang2018b, Wang2022a, Wang2022b, Dimauro2022, Kamieneski2023}. Our findings support an outside-in growth mechanism and imply that galaxies are shrinking in size due solely to in-situ star formation. This, in turn, suggests that in-situ star formation alone cannot fully explain the subsequent size growth of galaxies, challenging our current understanding of galaxy evolution beyond cosmic noon (Song et al. submitted).

Notably, our galaxy sample—especially those with lower stellar masses—exhibit higher central sSFR and younger stellar ages, reinforcing the notion of vigorous star formation in their inner regions. These trends in Figure \ref{SDprofile} collectively support a compaction process in high-redshift star-forming galaxies, where the inward migration of gas increases the central density and triggers highly efficient and rapid star formation in the inner region of galaxies. The dominant mechanisms driving this compaction may be frequent major mergers and the accretion of counter-rotating tidal streams \citep[e.g.,][]{Conselice2006, Zolotov2015, Tacchella2016, Wang2018}.

\subsection{$\boldsymbol{\xi_{\mathrm{ion}}}$ and $\boldsymbol{f_{\mathrm{esc}}}$ Profiles}

To estimate the ionizing photon escape fraction, $f_{\mathrm{esc}}$, we adopt the relation from \citet{Chisholm2022}, which links $f_{\mathrm{esc}}$ with $\beta$ (derived from fitting power law at rest-frame range of 1300\AA\ to 1800\AA\ of the spectra):
\begin{equation}
f_{\text {esc}}=(1.3 \pm 0.6) \times 10^{-4} \times 10^{(-1.22 \pm 0.1) \beta},
\end{equation}
in this context, \citet{Chisholm2022} found a strong correlation between $\beta$ and $f_{\mathrm{esc}}$ with a significance of 5.7$\sigma$.

Previous studies have shown that EW([OIII]) correlates strongly with various galaxy properties, such as stellar mass, metallicity, ionization parameters, and $\xi_{\mathrm{ion}}$ \citep[e.g.,][]{Reddy2018, Begley2025}. Galaxies with higher EW([OIII]) typically exhibit lower stellar masses, lower metallicities, and higher ionization parameters. Motivated by the potential contribution of strong [OIII]+H$\beta$ emitters to the ionizing photon budget, we investigate $\xi_{\mathrm{ion}}$ across our samples, using the $\xi_{\mathrm{ion}}$-EW([OIII]$\lambda5007$) relation from \citet{Tang2019}:

\begin{equation}
\log _{10}\left(\xi_{\text {ion }}\right)=0.76 \times \log _{10}\left(\mathrm{EW}\left(\left[\mathrm{O}{\text {III}}\right] \lambda 5007\right)\right)+ 23.27,
\end{equation}
and
\begin{equation}
\mathrm{EW}\left(\left[\mathrm{O}{\mathrm{III}}\right] \lambda 5007\right)  = 0.67 \times \mathrm{EW}\left(\left[\mathrm{O}{\mathrm{III}}\right]+\mathrm{H} \beta\right).
\end{equation}.

\begin{figure*}[htpb!]
    \centering
    \includegraphics[width=1\linewidth]{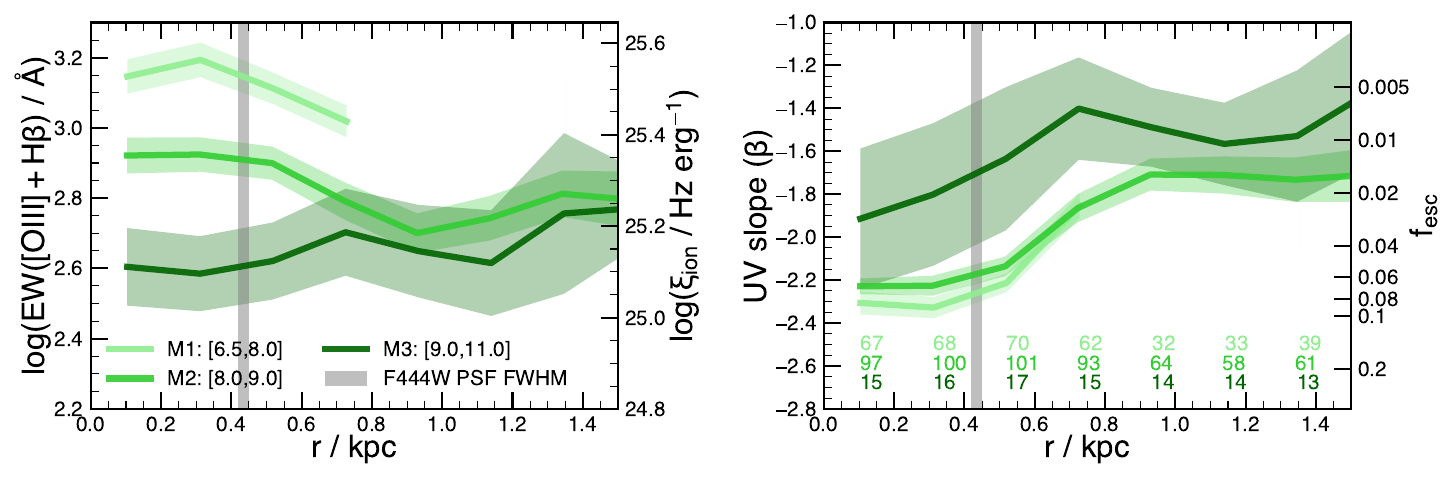}
    \caption{Radial profiles of [OIII]+H$\beta$ equivalent width and corresponding ionized photon production efficiency (left), and UV continuum slope and corresponding ionized photon escape fraction (right) of our galaxy sample. The color scheme, line styles, and shaded regions follow those in Figure \ref{SDprofile}. }
    \label{EWUVprofile}
\end{figure*}

Figure \ref{EWUVprofile} presents radial profiles of galaxy properties derived from SED fitting and ionizing photon diagnostics. The left panel shows the radial distribution of the [OIII]+H$\beta$ equivalent width and the ionizing photon production efficiency, $\xi_{\mathrm{ion}}$, in logarithmic scale. For low- and median-mass galaxies, these profiles decline with increasing radius, and on average, the gradient is steepest in low-mass galaxies. Specifically, $\xi_{\mathrm{ion}}$ drops from $\sim 10^{25.6}$ Hz erg$^{-1}$ at $r \sim$ 0.1 kpc to less than $\sim 10^{25.4}$ Hz erg$^{-1}$ in the outskirts. In contrast, the high-mass ones exhibit flat profiles. These results indicate that ionizing photons are produced with higher efficiency from the compact, actively star-forming centers of low-mass galaxies.

The right panel shows the UV continuum slope ($\beta$) and the escape fraction of ionizing photons ($f_{\mathrm{esc}}$). The $\beta$ profile steepens (i.e., becomes more negative) toward the galaxy center, while $f_{\mathrm{esc}}$ decreases with radius. In the lowest-mass galaxies, $\beta$ reaches as low as $-2.3$ and $f_{\mathrm{esc}}\sim$ 0.08 in the inner regions at $r \sim$ 0.1 kpc, but these values change to $\beta \sim$ $-1.7$ and $f_{\mathrm{esc}} \sim 0.02$ at the outer region. These patterns further support the scenario that more ionizing photons escape from the youngest and most centrally concentrated star-forming regions in low-mass galaxies.

Our derived properties and trends suggest that not all galaxies and radial regions at high redshift contribute equally to the ionizing photon budget. Instead, the central regions of galaxies have both higher ionized photon production efficiency and higher escape fraction, especially for low-mass galaxies, highlighting their potential role as important contributors to cosmic reionization. These trends are consistent with previous works on integrated galaxy properties in this redshift range. \citet{Begley2025} found a significant fraction of the lower-mass SFG population ($\log(M_*/\mathrm{M_\odot}) \leq 8.0$) display extreme optical emission-line equivalent widths (EW([OIII]+H$\beta$) $\geq$ 1000\AA). Such galaxies likely have lower metallicities, relatively bluer UV slopes, and are undergoing an intense surge in star formation, all of which become more common in the burstier population at these redshifts. Moreover, by examining Figures \ref{SDprofile} and \ref{EWUVprofile}, we find that the production and escape of ionizing photons are closely related to the presence of compact star-forming regions and high SFR surface densities within galaxies at $z\sim 7$. This finding is consistent with the results of \citet{Mascia2025}, who demonstrated that the escape of LyC photons from galaxies is primarily driven by internal mechanisms, such as compact star-forming regions and high SFR surface densities, rather than external mechanisms like mergers. This conclusion is also supported by low-redshift studies, such as the Lyman-alpha and Continuum Origins Survey (LaCOS), which found that escaping LyC emission is likely to originate from small, compact, and bright star-forming regions that create favorable conditions for increased stellar feedback \citep[e.g.,][]{LeReste2025}.

Despite the clear trends observed, our inferred values of $\xi_{\mathrm{ion}}$ and $f_{\mathrm{esc}}$ are subject to substantial uncertainties. The derivation of $\xi_{\mathrm{ion}}$ depends strongly on the assumptions embedded in stellar population synthesis models, including metallicity, the initial mass function, and the inclusion of binary stellar evolution \citep[e.g.,][]{Robertson2013, Robertson2015, Eldridge2017, Stanway2018, Shivaei2018, Seeyave2023}. Likewise, estimates of $f_{\mathrm{esc}}$ vary widely across methodologies and often rely on multivariate diagnostics involving several galaxy properties \citep[e.g.,][]{Chisholm2022, Flury2022, Jaskot2024a, Jaskot2024b, Mascia2025}. These strong model dependencies highlight the need for caution when interpreting absolute values of these parameters. Recently, \citet{Papovich2025} analyzed galaxies at $4.5 < z < 9.0$, and found the inferred $f_{\mathrm{esc}}$ remain generally low, with median values of $f_{\mathrm{esc}} < 1$–$3\%$, contrast strongly with expectations from scaling relations observed in low-redshift samples \citep[e.g.,][]{Flury2022} but consistent with predictions from simulations. 

It is important to emphasize that, as this work primarily focuses on exploring the relative radial profiles of ionization-related parameters, the simplified relations used to estimate $\xi_{\mathrm{ion}}$ and $f_{\mathrm{esc}}$ may not yield precise absolute values. Moreover, our derived values of $\xi_{\mathrm{ion}}$ and $f_{\mathrm{esc}}$ at $z \sim 7$ may also be biased by sample selection effects, particularly due to strong emission line and burstier star-formation histories.

\subsection{Total and Resolved Star Formation History}

\begin{figure*}[hptb]
    \centering
    \includegraphics[width=0.48\linewidth]{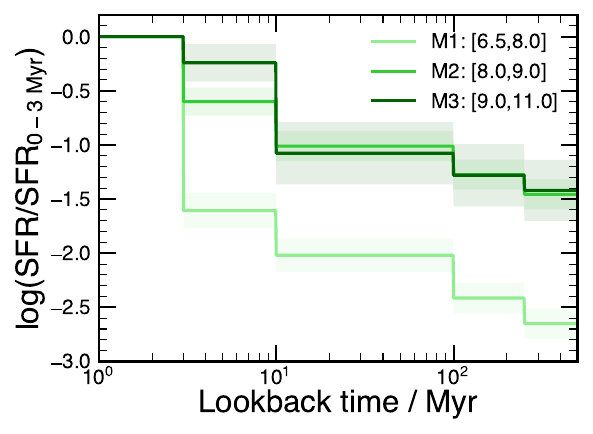}
    \includegraphics[width=0.48\linewidth]{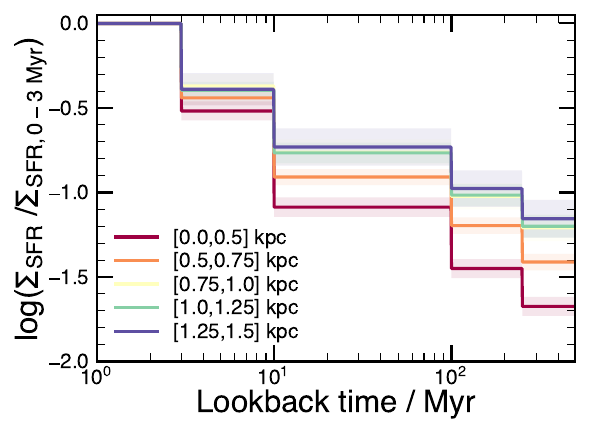}
    \caption{Normalized non-parametric ‘continuity’ star formation histories for each sample (left), and for each annulus across the entire sample (right). In both panels, the SFR and the SFR surface density (i.e., resolved SFR) are normalized to their values at the most recent time bin (0–3 Myr). Resolved SFHs for each sample in individual stellar mass bins are presented in Figure \ref{SFHprofile_Mbin} in the Appendix.}
    \label{SFHtotal}
\end{figure*}

Figure \ref{SFHtotal} presents the total and spatially resolved SFHs derived from SED fitting, normalized to their values in the most recent time bin (0–3 Myr). The left panel shows the total SFH for each galaxy sample, while the right panel displays the median $\Sigma_{\mathrm{SFR}}$ histories across different radial bins, averaged over the full sample. Detailed resolved SFHs for different galaxy samples are provided in the Appendix \ref{app3}. Overall, these galaxies statistically exhibit rising SFHs \citep{Whitler2023, Tacchella2023} and resolved SFHs at $z \sim 7$. This finding further supports our reasonable choice to employ a non-parametric SFH, which is well-suited to capture the complexities of their star formation processes.

For the total SFH, our galaxy samples display lower star formation activity at earlier times, followed by a pronounced increase in SFR in the most recent few Myr. This bursty behavior is especially evident in the low-mass galaxies, suggesting a recent episode of intense star formation.

For the resolved SFH, the innermost region ([0.0–0.5] kpc, purple) consistently shows the highest recent $\Sigma_{\mathrm{SFR}}$, which declines with increasing radius (e.g., [1.25–1.5] kpc, blue), where the outer regions formed earlier than the centers. This radial gradient in star formation activity reinforces the scenario of centrally concentrated, recent starbursts, particularly in low-mass galaxies. The combination of bursty central SFH and declining radial profiles may suggest a mode of star formation consistent with outside-in growth and central gas compaction for epochs prior to $z\sim 7$ \citep[e.g.,][]{Conselice2006, Zolotov2015, Tacchella2016, Wang2018}. 

\subsection{Limitation of photometric fitting of extremely emission line galaxies}

Photometric fitting of extreme emission line galaxies (EELGs) faces several limitations that may affect the accuracy of derived physical properties.

First, photometric redshift uncertainties may propagate into SED fitting results. We have also tested broader redshift priors, such as Gaussian distributions with $\mu_z=z_{\text{phot}}$ and $ \sigma_z=0.1,0.2,0.4$, and find that the resulting properties are generally stable within uncertainties. In addition, medium-band photometries (e.g., F410M) significantly enhance the accuracy of photometric redshifts by helping disentangle stellar continuum and nebular emission \citep{Harvey2025} and are effective in increasing completeness and reducing contamination by targeting strong emission lines within narrower redshift intervals \citet{Adams2025}.

Second, inaccurate modeling of nebular emission may bias SED results. Our configuration follows \citet{Begley2025}, who used BAGPIPES on $z>6$ galaxies with extreme [OIII]+H$\beta$ emission and employed fine (3 Myr) SFH time bins to resolve recent bursts. This approach allows accurate recovery of EWs in the presence of strong nebular features. In this work, the systematic uncertainties (estimated as $(P_{84} - P_{16}) / 2$) from SED fitting in individual regions for $\Sigma_*$, $\Sigma_{\mathrm{SFR}}$, mass-weighted-age, EW([OIII]+H$\beta$), and UV slope are approximately 0.16 dex, 0.16 dex, 59 Myr, 0.10 dex, and 0.25, respectively. Note that for most galaxy properties, the uncertainties in from SED fitting are generally much smaller than the observed radial variations. However, for stellar age, the SED fitting uncertainties are comparable to the size of the observed variation, which introduces greater uncertainty into the conclusions.

Third, the “outshining” effect can lead to underestimated stellar masses \citep{Harvey2025}. In low-mass galaxies with high specific SFRs and bursty SFHs, light from young OB stars may obscure older stellar populations. \citet{Harvey2025} showed that this can result in mass underestimates of $\sim$ 0.5 dex.

Fourth, photometric estimates of emission line strengths may differ from spectroscopic measurements. \citet{Duan2024} found that BAGPIPES-based photometric estimates generally agree with spectroscopic EWs, though color-based methods tend to underestimate EWs by $\sim 30\pm20\%$, likely due to continuum overestimation. Notably, galaxies with strong lines captured in medium bands (e.g., F410M) yield more reliable photometric EWs than those relying on broad bands.

In addition, we have tested performing SED fitting on the stacked images (see Appendix \ref{app1}) and repeating the earlier analyses, including radial profiles of galaxy properties and star formation histories. The similar trends shown in \ref{SDprofile}, Figure \ref{EWUVprofile}, and Figure \ref{SFHtotal} are recovered, confirming the robustness of our results.

In summary, while several systematic effects remain, we have taken steps to mitigate them through model configuration, sample selection, and error budgeting. We therefore regard our methodology as robust for deriving relative trends in galaxy properties, while interpreting absolute values with appropriate caution.

\section{Conclusions and Prospects} \label{sec:summary}

In this work, we present a spatially resolved analysis of 189 galaxies at $z \sim 6.7$–$7.6$ in the JADES GOODS-N and GOODS-S fields, selected using the JWST/NIRCam F410M filter. Unlike previous work that primarily examined integrated properties, we instead employ annular photometry measurements and SED fitting from rest-frame UV to optical wavelengths. This methodology allows us to investigate the internal star formation structures, spatially resolved star formation histories, ionizing photon production and escape, and their broader implications for cosmic reionization. Our main conclusions are as follows:

(1) The galaxies at $6.7 < z < 7.6$ occupy a relatively tight star formation main sequence (SFMS). The majority of these galaxies have stellar masses that are typically below $10^{9.5}\mathrm{M_\odot}$ (Figure~\ref{SFMS}). Their UV continuum slope is positively correlated with stellar mass, while the equivalent width of the [OIII]+H$\beta$ emission line is negatively correlated with stellar mass. Specifically, these galaxies display a median UV continuum slope of $\sim-2.07$ and a median EW([OIII]+H$\beta$) of $\sim$ 1000\AA\ (Figure~\ref{UVEW}).

(2) The radial profiles of stellar mass, SFR, sSFR, and mass-weighted age reveal that the galaxies, particularly those with lower stellar masses, host compact, centrally concentrated star-forming regions that are young and vigorously active (Figure \ref{SDprofile}). These observations are consistent with an outside-in growth scenario and may suggest that a compaction process is occurring at $z \sim 7$. This finding implies that galaxies could be shrinking in size primarily due to in-situ star formation, which challenges our current understanding of galaxy growth at high redshift.

(3) The compact centers of the galaxies exhibit extremely high EW([OIII]+H$\beta$)  (exceeding 1000\AA), elevated ionizing photon production efficiencies ($\xi_{\mathrm{ion}} \sim 10^{25.6}\ \mathrm{Hz\ erg}^{-1}$), steep UV slopes ($\beta \sim$ 2.3), and escape fractions above $f_{\mathrm{esc}} > 0.08$. These properties peak in the central region of galaxies and decline with radius, underscoring the important role of compact galaxies in driving ionizing photon output during the epoch of reionization, especially for the low-mass galaxies. In contrast, the high-mass ones exhibit relatively flat EW([OIII]+H$\beta$) profiles (Figure \ref{EWUVprofile}).

(4) The galaxies at $6.7 < z < 7.6$ exhibit rising star formation histories, characterized by intense, centrally concentrated bursts within the recent few Myr, especially in low-mass systems. (Figure \ref{SFHtotal}).

Taken together, our results highlight the pivotal role of the galaxies with strong [OIII] emission as key contributors to cosmic reionization. Their centrally concentrated, bursty star formation and efficient ionizing photon production and escape make them promising analogs for identifying the primary sources of cosmic reionization. Looking forward, future high-redshift observations with integral field unit (IFU) spectroscopy at finer spatial resolution will enable more detailed mapping of internal galaxy properties. Coupled with statistically larger samples, these advances will provide deeper insights into the nature of bursty star formation and the mechanisms of ionizing photon production and escape, ultimately enhancing our understanding of the role galaxies play in cosmic reionization.

These galaxies with strong [OIII]$\lambda$5007 emission may represent an early and extreme phase of galaxy evolution, analogous to local green peas but much smaller and more intense. Whether these galaxies are isolated phenomena or part of a continuous evolutionary sequence connecting the strong [OIII] emission systems and bright AGNs remains an open question. If their central black holes grow significantly alongside vigorous star formation, they may eventually evolve into luminous AGNs, suggesting a temporal sequence in galaxy evolution. Exploring the potential connection between these strong [OIII] emission systems and “Little Red Dots” \citep[e.g.,][]{Matthee2024} at high redshift—particularly in terms of gas concentration, intense starburst activity, and black hole accretion—will also be a valuable direction for future studies. Ultimately, understanding their fate is essential for revealing the early growth of supermassive black holes and the role of starbursts in shaping galaxy evolution.

\section{Acknowledgments}.

\begin{acknowledgments}
The authors thank the anonymous referee for their helpful comments that improved the quality of this Letter. EW thanks support of the National Science Foundation of China (Nos. 12473008) and the Start-up Fund of the University of Science and Technology of China (No. KY2030000200). The authors gratefully acknowledge the support of Cyrus Chun Ying Tang Foundations. 

The data products presented herein were retrieved from the Dawn JWST Archive (DJA). DJA is an initiative of the Cosmic Dawn Center (DAWN), which is funded by the Danish National Research Foundation under grant DNRF140.

\end{acknowledgments}

\vspace{5mm}

\appendix

\section{Stacked cutout images and surface brightness profiles}\label{app1}
\renewcommand{\thefigure}{A\arabic{figure}}
\setcounter{figure}{0} 

\begin{figure*}[hptb]
    \centering
    \includegraphics[width=0.60\linewidth]{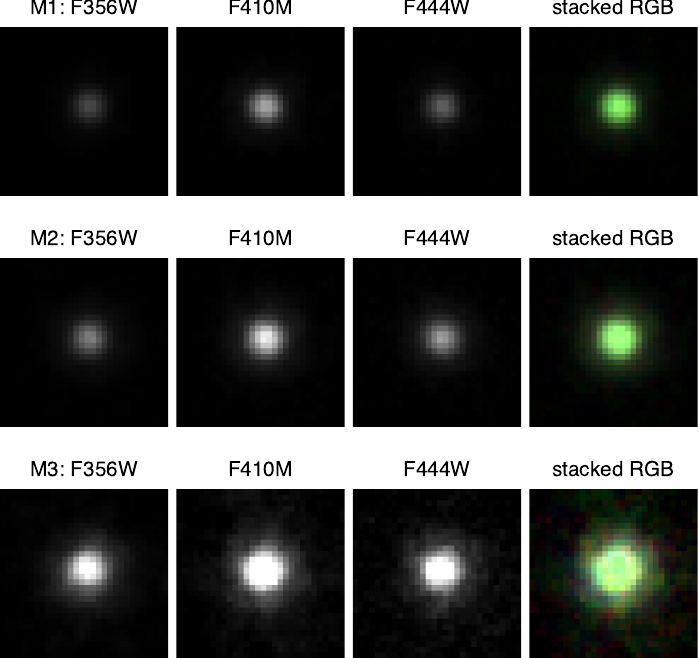}
    \includegraphics[width=0.36\linewidth]{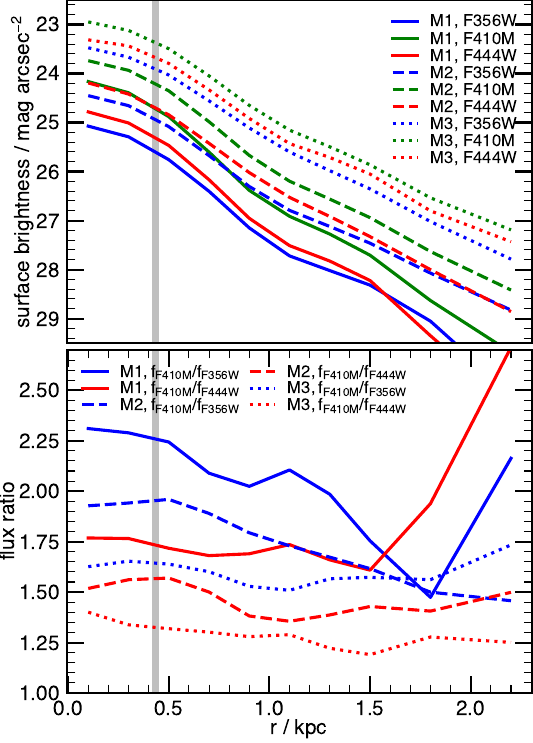}
    \caption{Left twelve panels: Stacked JWST NIRCam imaging cutouts of the galaxy sample of three stellar mass bins. From left to right, the panels show the F356W, F410M, and F444W bands, followed by an RGB composite image. The physical resolution of the cutouts is 0.2 kpc/pixel. Rightmost panels: Radial surface brightness profiles (top) and flux ratio profiles (bottom) of stacked galaxies. The profiles for the F356W, F410M, and F444W bands are shown in blue, green, and red, respectively. The surface brightness has been calculated by assuming the average kpc to arcsec scale at $<z> \sim 7.1166$ of our sample. The grey shaded area indicates the range of the F444W PSF FWHM across the redshift range of our sample.}
    \label{stack}
\end{figure*}

We resample the galaxy cutout images to a uniform physical resolution (0.2 kpc/pixel) and perform stacking separately for the galaxy samples of three stellar mass bins. As illustrated by the stacked images and surface brightness profiles in Figure\ref{stack}, a prominent feature is the flux excess at small radii, which is most evident in the F410M band, particularly for the low stellar mass (M1) bin. This enhancement indicates a strong central concentration of [OIII] emission in these galaxies.

\section{Radial Profiles of Galaxy Properties Normalized by Effective Radius}\label{app2}

\renewcommand{\thefigure}{B\arabic{figure}}
\setcounter{figure}{0}

\begin{figure*}[hptb]
    \centering
    \includegraphics[width=1\linewidth]{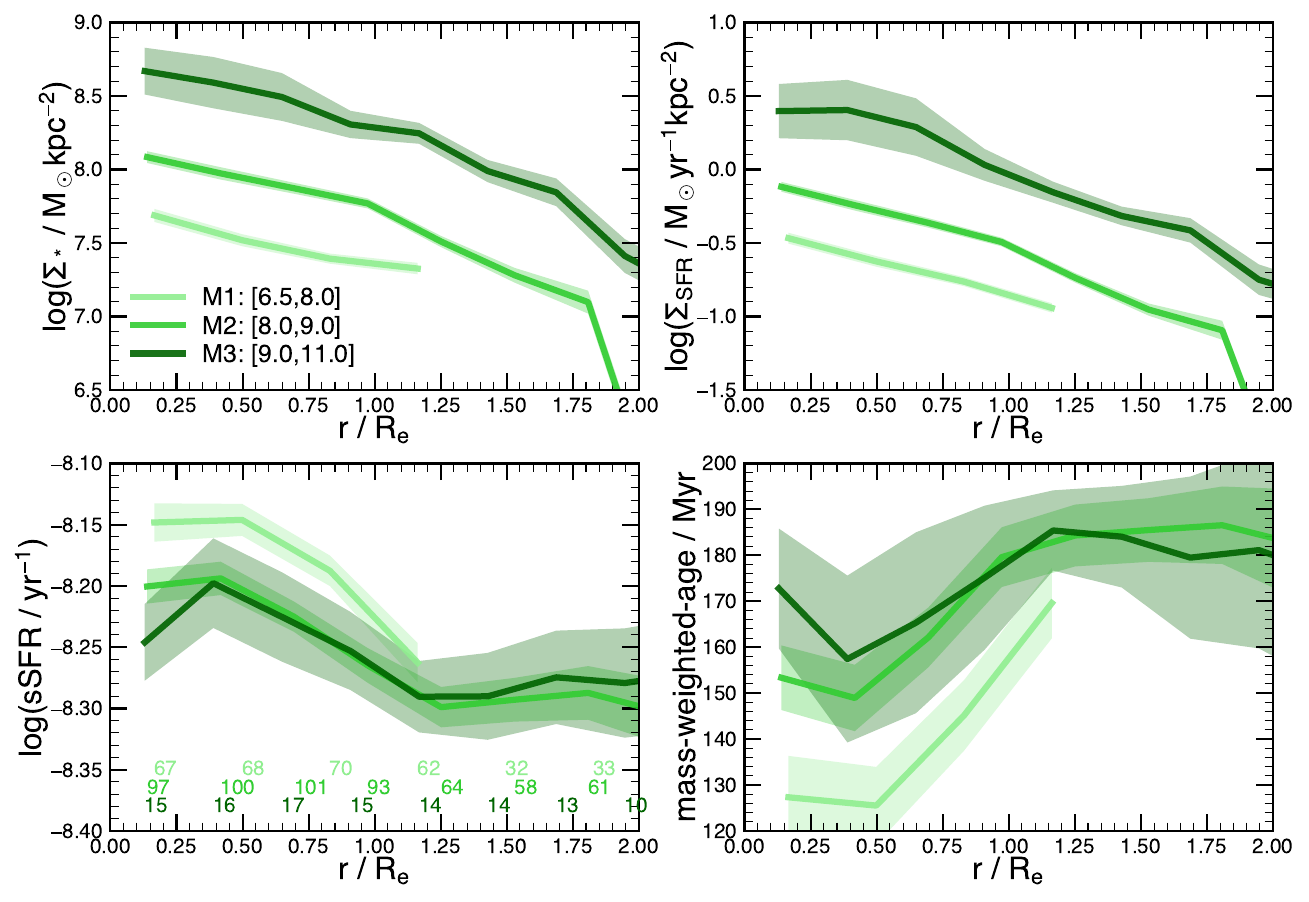}
    \caption{Similar to Figure \ref{SDprofile} but normalized by effective radius in x-axis.}
    \label{SDprofile_Re}
\end{figure*}

\begin{figure*}[hptb]
    \centering
    \includegraphics[width=1\linewidth]{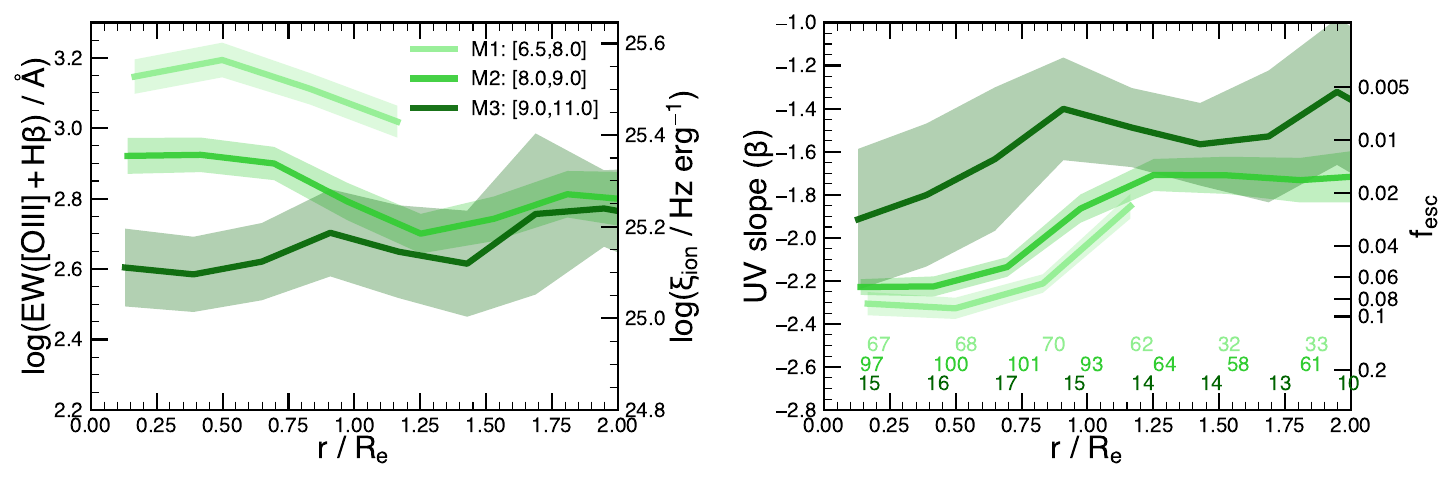}
    \caption{Similar to Figure \ref{EWUVprofile} but normalized by effective radius in x-axis.}
    \label{EWUVprofile_combined_Re}
\end{figure*}

Given that the stellar masses the galaxy samples span over three orders of magnitude, their physical sizes vary significantly. To account for this, we replot Figures \ref{SDprofile} and \ref{EWUVprofile} with the radial distance normalized by effective radii ($R_{\mathrm{e}}$), as shown in Figures \ref{SDprofile_Re} and \ref{EWUVprofile_combined_Re}, respectively.

At $z\sim 7$, individual galaxies often exhibit irregular and clumpy morphologies, making it challenging to measure reliable $R_{\mathrm{e}}$ on a per-galaxy basis. Therefore, we adopt the $R_{\mathrm{e}}$ values derived from stacked images for three stellar mass bins. Specifically, we resample each cutout image and stack them for each band, as done in Appendix~\ref{app1}. We then measure the half-light radius in each band and interpolate these measurements to estimate $R_{\mathrm{e}}$ at the rest-frame wavelength of 0.45 $\mu$m. The resulting $R_{\mathrm{e}}$ values for the three stellar mass bins ($\log(M_*/\mathrm{M}_\odot) = 6.5$–$8.0$ (M1), $8.0$–$9.0$ (M2), and $9.0$–$11.0$ (M3)) are approximately 0.626, 0.742, and 0.798 kpc, respectively. These values are consistent with those reported in \citet{Jia2024}.

As shown in the normalized profiles, the radial coverage extends beyond $1R_{\mathrm{e}}$ and up to $2R_{\mathrm{e}}$ in some cases. Compared to the original profiles plotted in physical (kpc) scale, the overall trends remain qualitatively unchanged, indicating that the structural features are robust to this normalization.

\section{Star Formation History Profiles for Different Galaxy Samples}\label{app3}

\renewcommand{\thefigure}{C\arabic{figure}} 
\setcounter{figure}{0} 

\begin{figure*}[hptb]
    \centering
    \includegraphics[width=1\linewidth]{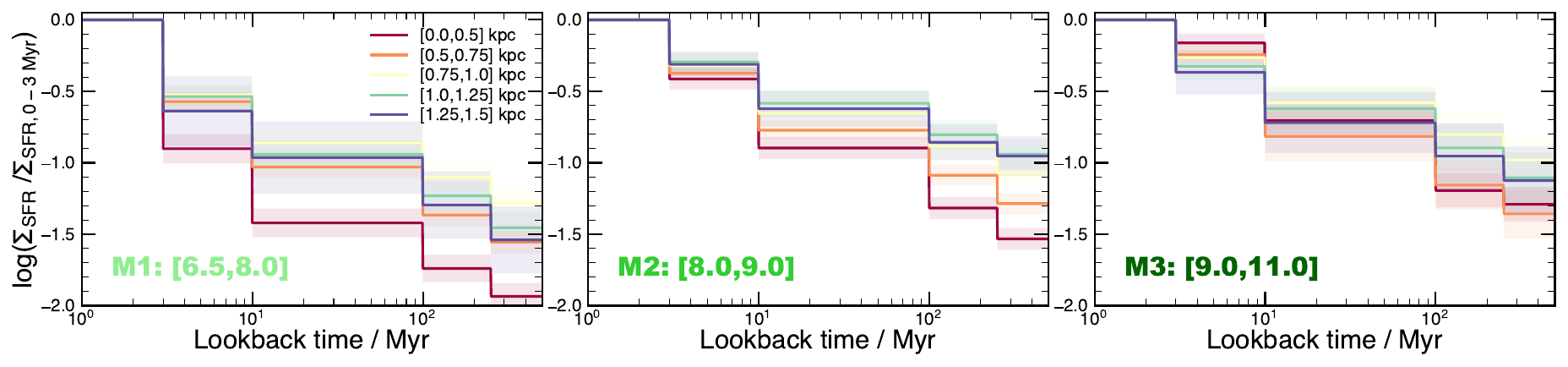}
    \caption{Similar to the right panel of Figure \ref{SFHtotal} but shown separately for three distinct samples of different stellar mass, as indicated in the bottom-left corner of each panel.}
    \label{SFHprofile_Mbin}
\end{figure*}

Figure \ref{SFHprofile_Mbin} shows the resolved SFH ($\Sigma_{\mathrm{SFR}}$ as a function of lookback time) for three galaxy samples of different stellar mass. On average, the innermost region ([0.0–0.5] kpc, purple) consistently shows the highest recent $\Sigma_{\mathrm{SFR}}$, which declines with increasing radius (e.g., [1.25–1.5] kpc, blue). This trend is consistent with that observed in the right panel of Figure \ref{SFHtotal}. In addition, low-mass galaxies tend to exhibit more recently rising SFHs compared to their higher-mass counterparts.

\bibliography{sample631}{}
\bibliographystyle{aasjournal}
\end{document}